\documentclass[11pt]{article}
\usepackage{amsfonts}
\usepackage{amsmath}
\usepackage{amsthm}
\usepackage{amssymb}
\usepackage[mathscr]{eucal}
\usepackage{graphics}

\def\odotacc{\mathaccent "7017 }

\def\sR{{\Bbb R}}

\def\F{{\mathscr{F}}}

\def\N{{\mathscr{N}}}

\def\bfeta{\boldsymbol{\eta}}

\def\bfalpha{\boldsymbol{\alpha}}
\def\bfphi{\boldsymbol{\varphi}}
\def\int{\intop}

\newcommand{\wotwo}{\mathop{\odotacc{W}^{1,2}}}

\newcommand{\woq}{\mathop{\odotacc{W}^{k,q}}}

\def\div{\hbox{\rm div}\,}

\begin{document}
\title{On the flux problem in the theory of steady
Navier--Stokes equations with nonhomogeneous boundary
conditions\footnote{
bounded domains,  Stokes system, stationary Navier Stokes equations,
boundary--value problem.}}
\author{ Mikhail V. Korobkov\footnote{Sobolev Institute of Mathematics, Acad. Koptyug pr. 4,
630090 Novosibirsk, Russia; korob@math.nsc.ru},  Konstantin
Pileckas\footnote{Faculty of Mathematics and Informatics, Vilnius
University, Naugarduko Str., 24, Vilnius, 03225  Lithuania;
pileckas@ktl.mii.lt} \, and Remigio Russo\footnote{ Dipartimento di
Matematica, Seconda Universit\`a di Napoli,  via Vivaldi 43, 81100
Caserta, Italy; remigio.russo@unina2.it}$\;$}

 \date{}
\maketitle

\begin{abstract} We study the nonhomogeneous boundary value problem
for Navier--Stokes equations of steady motion of a viscous
incompressible fluid in  a two--dimensional   bounded multiply
connected domain $\Omega=\Omega_1\setminus\overline{\Omega}_2,
\;\overline\Omega_2\subset \Omega_1$. We prove that this problem has
a solution if the flux $\F$ of  the boundary datum through
$\partial\Omega_2$ is nonnegative (outflow condition).
\end{abstract}

\bigskip
\section{Introduction}
Let $\Omega$ be a bounded multiply connected domain in $\sR^n,\,
n=2,3$, with Lipschitz boundary $\partial\Omega$ consisting of $N$
disjoint components $\Gamma_j$, i.e.
$\partial\Omega=\Gamma_1\cup\ldots\cup\Gamma_N$ and
$\Gamma_i\cap\Gamma_j=\emptyset,\, i\neq j$. In $\Omega$ consider
the stationary  Navier--Stokes system  with nonhomogeneous
boundary conditions
\begin{displaymath} 
\left\{ \begin{array}{rcl} 
-\nu \Delta{\bf u}+\big({\bf u}\cdot \nabla\big){\bf u} +\nabla p & = & {0}\qquad \hbox{\rm in }\;\;\Omega,\\[4pt]
\div\,{\bf u} &  = & 0  \qquad \hbox{\rm in }\;\;\Omega,\\[4pt]
 {\bf u} &  = & {\bf a} \qquad \hbox{\rm on }\;\;\partial\Omega.
 \end{array}\right.
 \eqno(1)
\end{displaymath}

Starting from the famous J. Leray's paper \cite{Leray} published in
1933, problem (1) was a subject of investigation in many papers
\cite{Amick}, \cite{BOPI}--\cite{Galdibook},
\cite{Hopf}--\cite{Lad}, \cite{Lions}, \cite{Takashita},
\cite{Temam}, \cite{VorJud}. The continuity equation $(1_2)$ implies
the necessary compatibility condition for the solvability of problem
(1):
$$
\int\limits_{\partial\Omega}{\bf a}\cdot{\bf
n}\,dS=\sum\limits_{j=1}^N\int\limits_{\Gamma_j}{\bf a}\cdot{\bf
n}\,dS=0,\eqno(2)
$$
where ${\bf n}$ is a unit vector of the outward (with respect to
$\Omega$) normal to $\partial\Omega$. However, for a long time the
existence of a weak solution ${\bf u}\in W^{1,2}(\Omega)$  to
problem (1) was proved only under the condition
$$
{\F}_j=\int\limits_{\Gamma_j}{\bf a}\cdot{\bf n}\,dS=0,\qquad
j=1,2,\ldots,N, \eqno(3)
$$
(see  \cite{Leray}, \cite{Lad1}, \cite{Fu}, \cite{VorJud}
\cite{Lad}, etc.). Condition (3) requires the fluxes ${\F}_j$ of the
boundary datum ${\bf a}$ to be zero separately on all components
$\Gamma_j$ of the boundary $\partial\Omega$, while the compatibility
condition (2) means only that the total flux is zero. Thus, (3) is
stronger than (2) and (3) does not allow the presence of sinks and
sources.

Problem (1), (3) was first studied by J. Leray \cite{Leray} who
initiated two different approaches to prove its solvability. In both
approaches the problem is reduced to an operator equation with  a
compact operator and the existence of a fixed--point is obtained by
using the Leray--Schauder theorem. The main difference in these
approaches is in getting an a priori estimate of the solution. The
first method uses the extension of boundary data ${\bf a}$ into
$\Omega$ as ${\bf A}(\varepsilon,
x)={curl}\,\big(\zeta(\varepsilon,x) {\bf b}(x)\big)$, where
$\zeta(\varepsilon,x)$   is Hopf's cut--off function \cite{Hopf}.
For such extension there holds an estimate (see, e.g., \cite{Lad})
$$
\Big|\int\limits_\Omega \big({\bf v}\cdot\nabla\big){\bf A}\cdot{\bf
v}\,dx\big|\leq \varepsilon c \int\limits_\Omega |\nabla{\bf
v}|^2\,dx\quad\forall\;\; {\bf v}\in\wotwo(\Omega),\eqno(4)
$$
with $c$ being independent of $\varepsilon$ and $\varepsilon >0$
taken sufficiently  small (so that $\varepsilon c<\nu$). Obviously,
the extension of the boundary data in the form of $curl$ is possible
only if condition (3) is satisfied. A. Takashita \cite{Takashita}
has constructed a counterexample showing that estimate (4) is false
whatever the choice of the extension ${\bf A}$ can be, if the
condition (3) is not valid. Thus, the first approach may be applied
only when (3) is valid.

The second approach is to prove an a priory estimate by
contradiction. Such arguments also could be found in the book of
O.A. Ladyzhenskaya \cite{Lad}. Later, a slight modification of this
argument was proposed independently by L.V. Kapitanskii and K.
Pileckas \cite{KaPi1}, and by Ch.J. Amick \cite{Amick}. This
modification has the advantage that it allows to take any solenoidal
extension of the boundary data and requires (unlike Hopf's
construction) only the Lipschitz regularity of the boundary
$\partial\Omega$. We should mention that the method used in
\cite{Amick}, \cite{KaPi1} was already contained in the basic paper
of J.Leray \cite{Leray}. In \cite{KaPi1} the solvability of problem
(1) was proved by this method only under "stronger" condition (3),
while in \cite{Amick} was constructed a class of plane domains with
special symmetry on $\Omega$ and on ${\bf a}=\big(a_1, a_2\big)$,
where problem (1) is solvable for arbitrary fluxes ${\F}_j$,
assuming only condition (2). More precisely, it is proved in
\cite{Amick} that problem (1) has at least one solution for all
values of ${\F}_j$, if $\Omega\subset \sR^2$ is symmetric with
respect to the $x_1$--axis and all components $\Gamma_j$ intersect
the line $\{x: \;x_2=0\}$, ${a}_1$ is an even function, while
${a}_2$ is an odd function with respect to $x_2$. Note that Amick's
result was proved by contradiction and does not contain an effective
a priori estimate for the  Dirichlet integral of the solution. An
effective estimate for the solution of the Navier--Stokes problem
with the above symmetry conditions was first obtained by H. Fujita
\cite{Fu1} (see also \cite{Morimoto}). Recently V.V. Pukhnachev has
established  an analogous estimate for the solution to problem (1)
in the case of three--dimensional stationary fluid motion with two
mutually perpendicular planes of symmetry (private communication).

The  assumption  on ${\F}_j$ to be zero (see (3)) was relaxed in
\cite{Galdi1} where it is shown  that problem (1)  still admits a
solution  provided that $|{\F}_j|$ are sufficiently
small\footnote{As far as we are aware, the idea of  requiring
smallness of $|{\F}_j| $ instead of its vanishing appears for the
first time in \cite{Finn} (see also \cite{Fu}).}. In \cite{BOPI}
estimates for $|{\F}_j|$ are expressed in terms of simple geometric
characteristics of $\Omega$ which can be easily verified for
arbitrary domains. These results have been extended to solutions
corresponding to boundary data in Lebesgue's spaces in \cite{Russo}.
As far as exterior domains are concerned, the hypothesis of zero
flux at the boundary has been replaced by the assumptions of small
flux in \cite{RussoA}.

An interesting contribution to the Navier--Stokes problem  is due to
H.Fuji-ta and H. Morimoto  \cite{FM} (see also
\cite{RussoAStarita}). They studied problem (1) in a domain $\Omega$
with two components of the boundary $\Gamma_1$ and $\Gamma_2$.
Assuming that ${\bf a}={\F} \nabla u_0+\bfalpha$, where ${\F}\in{
\sR}$, $u_0$  is a harmonic function, and $\bfalpha$ satisfies
condition  (3),  they proved that there is a countable subset
$\N\subset {\sR}$ such that if ${\F}\not\in\N$ and $\bfalpha$ is
small (in a suitable norm), then system (1) has a weak solution.
Moreover, if $\Omega\subset \sR^2$ is an annulus and $u_0=\log|x|$,
then $\N=\emptyset$.

To the best of our knowledge this is  the state of art of the
Navier--Stokes problem with nonhomogeneous boundary conditions in
bounded multiply connected domains. As a consequence, the
fundamental question whether problem (1) is solvable for all
values of ${\F}_j$ (Leray's problem) is still open despite of
efforts of many mathematicians.

In this paper we study problem (1) in a plane domain
$$
\Omega= \Omega_1\setminus\overline{\Omega}_2, \quad
\overline{\Omega}_2\subset\Omega_1, \eqno(5)
$$
where $\Omega_1$ and $\Omega_2$ are bounded simply connected domains
of $\sR^2$ with Lip-schitz boundaries $\partial \Omega_1=\Gamma_1$,
$\partial \Omega_2=\Gamma_2$. Without loss of generality we may
assume that $\Omega_2\supset \{x\in\sR^2: |x|<1\}$.  Since $\Omega$
has only two components of the boundary, condition (2) may be
rewritten in the form
$$
\F=\int\limits_{\Gamma_2}{\bf a}\cdot{\bf
n}\,dS=-\int\limits_{\Gamma_1}{\bf a}\cdot{\bf n}\,dS\eqno(6)
$$
(${\bf n}$ is an outward  normal with respect to the domain
$\Omega$).  Using some suggestions from  \cite{Amick}, we prove that
problem (1) is solvable without any restriction on the value of
$|\F|$ provided ${\F}\ge 0$ (outflow condition). Note that this is
the first result on Leray's problem which does not require smallness
or symmetry conditions of the data.

This results was first announced in the "International Conference on
Mathematical Fluid Mechanics: a Tribute to Giovanni Paolo Galdi",
May 21-25, 2007, Portugal
(http://cemat.ist.utl.pt/gpgaldi/abs/russo.pdf).

\section{Notation and preliminary results}

Everywhere in the paper
$\Omega=\Omega_1\setminus\overline{\Omega}_2\subset\sR^2$ is a
bounded domain defined above by (5). We assume that the boundary
$\partial\Omega$ is Lipschitz  \footnote{$\partial\Omega$ is
Lipschitz, if for every $\xi\in\partial\Omega$, there is a
neighborhood of $\xi$ in which $\partial\Omega$  is the graph of a
Lipschitz continuous function (defined on an open interval).}. We
use standard notations for function spaces: $C(\overline\Omega)$,
$C(\partial\Omega)$, $W^{k,q}(\Omega)$, $\woq(\Omega)$,
$W^{\alpha,q}(\partial\Omega)$, where $\alpha\in(0,1), k\in{\Bbb
N}_0, q\in[1,+\infty]$. ${\cal H}^1(\sR^2)$ denotes the Hardy space
on ${\Bbb R}^2$. In our notation we do not distinguish function
spaces for scalar and vector valued functions; it is clear from the
context whether we use scalar or vector (or tensor) valued function
spaces. $H(\Omega)$ is subspace of all divergence free vector fields
from $\wotwo(\Omega)$ with the norm
$$
\|{\bf u}\|_{H(\Omega)}=\|\nabla{\bf u}\|_{L^2(\Omega)}.
$$
Note that for function ${\bf u}\in H(\Omega)$ the norm
$\|\,\cdot\,\|_{H(\Omega)}$ is equivalent to
$\|\,\cdot\,\|_{W^{1,2}(\Omega)}$.\\

 Let us collect auxiliary results that we shall
use below to prove the solva-bility of problem (1).
\\

{\bf Lemma 1.} {\it Let $\Omega $ be a bounded domain with Lipschitz
boundary. If ${\bf a}\in W^{1/2,2}(\partial\Omega)$
and
$$
\int\limits_{\partial\Omega }{\bf a}\cdot{\bf n}\,dS=0,
$$
then there exists a divergence free extension ${\bf A}\in
W^{1,2}(\Omega)$ of ${\bf a}$ such that
$$
 \|{\bf
A}\|_{W^{1,2}(\Omega)}\leq c \|{\bf
a}\|_{W^{1/2,2}(\partial\Omega)}.\eqno(7)
$$}

Lemma 1 is well known (see \cite{LadSol1}).
\\

{\bf Lemma 2.} (see \cite{SolSca}). {\it Let  $\Omega $ be a bounded
domain with Lipschitz boundary and let $R(\bfeta)$
be a continuous linear functional defined on $\wotwo(\Omega)$. If
$$
R(\bfeta)=0\qquad\forall\;\;\bfeta\in H(\Omega),
$$
then there exists a function $p\in L^2(\Omega)$ with
$\int\limits_\Omega p(x)\,dx=0$ such that
$$
R(\bfeta)= \int\limits_\Omega p\,{\rm
div}\,\bfeta\,dx\qquad\forall\;\;\bfeta\in \wotwo(\Omega).
$$
Moreover, $\|p\|_{L^2(\Omega)}$ is equivalent to
$\|R\|_{(\wotwo(\Omega))^*}$. }
\\

{\bf Lemma 3.}  {\it Let $f\in {\cal H}^1(\sR^2)$ and let
$$
J(x)=\int_{{\Bbb R}^2}\log|x-y|\,f(y)\,dy.\eqno(8)
$$
Then \\

(i) $J\in C(\sR^2)$;\\

(ii)  $\nabla J\in L^{2}(\sR^2)$, $D^\alpha J\in L^{1}(\sR^2),
|\alpha|=2$. }
\\

Lemma 3 is well known; a proof of the property (i) could be found in
\cite{Tailorbook}  (see Theorem 5.12 and Corollary 12.12 at p.
82--83), and the property (ii) is proved, for example,  in
\cite{catafalco}
(see Theorem 5.13, p. 208).\\

{\bf Lemma 4.} {\it Let ${\bf w}\in W^{1,2}(\sR^2)$ and $\div{\bf
w}=0$. Then
$$
\div\big[\big({\bf w}\cdot\nabla\big){\bf
w}\big]=\sum\limits_{i,j=1}^2\frac{\partial  w_i}{\partial
x_j}\frac{\partial w_j}{\partial x_i}\in {\cal H}^1(\mathbb{R}^2).
$$
}\\

Lemma 4 follows from div-curl lemma with two cancelations (see,
e.g., Theorem II.1 in \cite{CLMS}).\\

{\bf Lemma 5.} {\it Let $\Omega\subset\sR^2$ be a bounded domain
with Lipschitz boundary  and let $h\in C(\partial\Omega)$. If $h$
could be extended into domain $\Omega$ as a function $H\in
W^{1,2}(\Omega)$, then there exists a unique  weak solution $v\in
W^{1,2}(\Omega)$ of the problem
\begin{displaymath} \left\{\begin{array}{rcl} 
 -\Delta{v} & = & 0\qquad \hbox{\rm in }\Omega,\\[4pt]
{v} & = & {h} \qquad \hbox{\rm on }\partial\Omega,
\end{array}\right.\eqno(9)
\end{displaymath}
such that $v\in C(\overline{\Omega})$.
 }
\\

The proof of Lemma 5 could be found in  \cite{Littman} (see also
Theorem 4.2 in \cite{Landis}). Note that not every continuous on
$\partial\Omega$ function $h$ could be extended into $\Omega $ as a
function $H$ from $W^{1,2}(\Omega)$. If  this is the  case, then
there exists a weak solution $v$ of (9) satisfying only $v\in
W_{loc}^{1,2}(\Omega)\cap C(\overline{\Omega})$ (see Chapter II in
\cite{Landis}).\\

\section{Euler equation}

In this section we collect some properties of a solution to the
Euler system
\begin{displaymath}
\left\{\begin{array}{rcl} \big({\bf
w}\cdot\nabla\big){\bf w}+\nabla p & = & 0,\\[4pt]
\div{\bf w} & = & 0,
\end{array}\right. \eqno(10)
\end{displaymath}
that are used below to prove the main result of the paper.
\\

Assume that   ${\bf w}\in W^{1,2}(\Omega)$ and $p\in
W^{1,2}(\Omega)$ satisfy  the Euler equations (10) for almost all
$x\in\Omega$ and let $\int\limits_{\Gamma_i}{\bf w}\cdot{\bf
n}dS=0,\;i=1,2$. Then there exists a continuous stream function
$\psi\in W^{2,2}(\Omega)$ such that $\nabla\psi=(-w_2, w_1)$. Denote
by  $\Phi=  p+\frac{|{\bf w}|^2}{2}$ the total head pressure
corresponding to the solution $({\bf w}, p)$. Obviously, $\Phi\in
W^{1,s}(\Omega)$ for all $s\in [1,2)$. By direct calculations one
can easily get the identity
$$
\nabla\Phi\equiv \Big(\frac{\partial w_2}{\partial
x_1}-\frac{\partial w_1}{\partial x_2}\Big)\big(w_2,
-w_1\big)=(\Delta \psi)\nabla\psi.\eqno(11)
$$
If all functions are smooth, from this identity the classical
Bernoulli  law follows immediately: {\it the total head pressure
$\Phi(x)$ is constant along any streamline of the flow}.

In the general case  the following assertion holds.
\\

{\bf Lemma 6.}\cite{korob1}. {\it Let ${\bf w}\in W^{1,2}(\Omega)$
and $p\in W^{1,2}(\Omega)$ satisfy  the Euler equations (10) for
almost all $x\in\Omega$ and let $\int\limits_{\Gamma_i}{\bf
w}\cdot{\bf n}dS=0,\;i=1,2$. Then for any connected set $K\subset
\overline\Omega$ such that
$$
\psi(x)\big|_{K}=const,\eqno(12)
$$
the identity
$$
\Phi(x)=const \quad   \mathfrak{H}^1-almost\quad everywhere\quad on
\quad K\eqno(13)
$$
holds. Here $\mathfrak{H}^1$ denotes one-dimensional Hausdorff
measure\footnote{$\mathfrak{H}^1(F)=\lim\limits_{t\to 0+}\mathfrak{H}^1_t(F)$, where
$\mathfrak{H}^1_t(F)=\inf\{\sum\limits_{i=1}^\infty {\rm diam}
F_i:\, {\rm diam} F_i\leq t, F\subset \bigcup\limits_{i=1}^\infty
F_i\}$. }.

In particular, if ${\bf w}=0$ on ${\partial \Omega}$ $($in the sense
of trace$)$, then the pressure $p(x)$ is constant on
$\partial\Omega$. Note that $p(x)$ could take different constant
values $p_j=p(x)\big|_{\Gamma_j}, j=1,2$,  on different components
$\Gamma_j$ of the boundary $\partial\Omega$. }\\

Here and henceforth we understand connectedness in the sense of
general topology. Note that the proof of the above lemma is based on
classical results of \cite{Kronrod} and on recent results obtained
in \cite{korob}. The last statement of Lemma 6  was proved in
\cite{KaPi1} (see Lemma 4) and in \cite{Amick} (see Theorem 2.2).
\\

{\bf Lemma 7.} {\it Let  $({\bf w}, p)$ satisfy  the Euler equations
(10) for almost all $x\in\Omega$, ${\bf w}\in W^{1,2}(\Omega)$ and
${\bf w}(x)\big|_{\partial\Omega}=0$. Then
$$
p\in C(\overline{\Omega})\cap W^{1,2}(\Omega).\eqno(14) $$}

{\bf Proof.} From Euler equations (10) it follows that $p\in
W^{1,s}(\Omega)$ for any $s\in[1,2)$ and
$$
\|p\|_{W^{1,s}(\Omega)}\leq c\|{\bf w}\|_{H(\Omega)}^2.
$$
  Multiply (10) by
$\bfphi=\nabla\xi$, where $\xi\in C_0^\infty(\Omega)$:
$$
\int_\Omega\nabla p\cdot\nabla\xi\,dx=-\int\limits_\Omega\big({\bf
w}\cdot\nabla\big){\bf w}\cdot\nabla\xi\,dx \quad\forall\xi\in
C^\infty_0(\Omega).
$$
Thus, $p\in W^{1,q}(\Omega)$ is the unique weak solution of the
boundary value problem for the Poisson equations
\begin{displaymath}
\left\{\begin{array}{rcl} -\Delta p & = & \div\big[\big({\bf
w}\cdot\nabla\big){\bf w}\big]\qquad \hbox{\rm in }\;\;\Omega,
\\[4pt]
p(x)& = & p_1 \qquad\qquad\qquad\quad\;\; \hbox{\rm on
}\;\,\Gamma_1,\\[4pt]
p(x)  & = &  p_2  \,\;\qquad\qquad\qquad\quad\; \hbox{\rm on
}\;\;\Gamma_2.
\end{array}\right.\eqno(15)
\end{displaymath}
According to  Lemma 4, $\div\big[\big({\bf w}\cdot\nabla\big){\bf
w}\big] \in {\cal H}^1(\sR^2)$ (here we assume that ${\bf w}\in
H(\Omega)$ is extended by zero to $\sR^2$). Define  the function
$J_1(x)$ by the formula
$$
J_1(x)=-\frac{1}{2\pi}\int_{{\sR}^2}\log|x-y|\,{\rm
div}_y\,\big[\big({\bf w}(y)\cdot\nabla_y\big){\bf w}(y)\big]dy.
$$
In virtue of Lemma 3,  $J_1\in C(\sR^2)$, $\nabla J_1\in
L^{2}({\sR}^2)$, $D^\alpha J_1\in L^{1}({\sR}^2), |\alpha|=2$. Since
$-\Delta J_1(x)= \div\big[\big({\bf w}\cdot\nabla\big){\bf w}\big]$
in $\sR^2$,  we get for $J_2(x)=p(x)-J_1(x)$ the following problem
\begin{displaymath}
\left\{\begin{array}{rcl} -\Delta  J_2 & = & 0\qquad\qquad\qquad
\hbox{\rm in }\;\;\Omega,
\\[4pt]
J_2\big|_{\partial\Omega} & = & j_2-j_1\;\;\qquad\;\;\;\; \hbox{\rm
on }\;\;\partial\Omega,
\end{array}\right.\eqno(16)
\end{displaymath}
where $j_1(x)=J_1(x)\big|_{\partial\Omega}$,
$$
j_2(x)=\left\{\begin{array}{rcl} {p}_1 \qquad\; \hbox{\rm on}\;\;\Gamma_1,\\[4pt]
{p}_2\qquad\; \hbox{\rm on }\;\;\Gamma_2.
\end{array}\right.
$$
The function $j_1$ is a trace on $\partial\Omega$ of $J_1\in
W^{1,2}(\Omega)\cap C(\overline{\Omega})$, while $j_2\in
C(\partial\Omega)$ and $j_2$ obviously could be extended to $\Omega$
as a function from $W^{1,2}(\Omega)$. Thus, by Lemma 5 problem (16)
has a unique weak solution $J_2\in W^{1,2}(\Omega)$ such that
$J_2\in C(\overline{\Omega})$. By uniqueness  $ p(x)=J_1(x)+J_2(x)$.
Hence, $p\in C(\overline{\Omega})\cap
W^{1,2}(\Omega)$.  \\

We say that the function $f\in W^{1,s}(\Omega)$  satisfies a {\it
weak one-side maximum principle locally} in $\Omega$, if
$$
\mathop{\hbox{\rm ess}\,\hbox{\rm sup}}_{x\in\Omega^\prime}\,f(x)\leq \mathop{\hbox{\rm
ess}\,\hbox{\rm sup}}_{x\in\partial\Omega^\prime}\,f(x) \eqno(17)
$$
holds for any strictly interior subdomain $\Omega^\prime$
($\overline{\Omega^\prime}\subset\Omega)$ with the boundary
$\partial\Omega^\prime$ that does not contain singleton connected
components. (In (17) negligible sets are the sets of 2--dimensional
Lebesgue measure zero in the left ess sup, and the sets of
1--dimensional Hausdorff measure  zero in the right ess sup.) If
(17) holds for any $\Omega^\prime \subset\Omega$ with the boundary
$\partial\Omega^\prime$ not containing singleton connected
components, then we say that $f\in W^{1,s}(\Omega)$
 satisfies a {\it weak  one-side maximum principle } in $\Omega$
(since the boundary $\partial\Omega$ is Lipschitz, we can take
$\Omega^\prime=\Omega$ in (17)).
\\

{\bf Lemma 8.} \cite{korob1}. {\it Let ${\bf w}\in W^{1,2}(\Omega)$
and $p\in W^{1,2}(\Omega)$ satisfy  the Euler equations (10) for
almost all $x\in\Omega$ and let $\int\limits_{\Gamma_i}{\bf
w}\cdot{\bf n}dS=0,\;i=1,2$. Assume that there exists a sequence of
functions $\{\Phi_\mu\}$ such that $\Phi_\mu\in
W^{1,s}_{loc}(\Omega)$ and $\Phi_\mu\rightharpoonup\Phi$ in the
space $W^{1,s}_{loc}(\Omega)$ for all $s\in (1,2)$. If all
$\Phi_\mu$  satisfy the weak one-side maximum principle locally in
$\Omega$, then $\Phi$ satisfies the weak one-side maximum principle
in $\Omega$.

In particular, if ${\bf w}\big|_{\partial\Omega}=0$, then
$$
\mathop{\hbox{\rm ess}\,\hbox{\rm sup}}_{x\in\Omega}\,\Phi(x)\leq
\mathop{\hbox{\rm ess}\,\hbox{\rm
sup}}_{x\in\partial\Omega}\,\Phi(x)= \max\{p_1, p_2\}.\eqno(18)
$$}

The proof of the above lemma is based on Lemma 6, classical results
of \cite{Kronrod}, and on recent results obtained in \cite{korob}.
Note that the weaker version of Lemma 8 was proved by Ch. Amick
\cite{Amick} (see Theorem 3.2 and Remark thereafter).
\\

\section{Existence theorem}

Let us consider Navier--Stokes problem (1) in the domain $\Omega$
defined by (5) and assume that $\partial\Omega$ is at least
Lipschitz. If the boundary datum ${\bf a}\in
W^{1/2,2}(\partial\Omega)$  and  ${\bf a}$ satisfies the condition
(6), i.e.,
$$
\int\limits_{\partial\Omega}{\bf a}\cdot{\bf
n}\,dS=\int\limits_{\Gamma_1}{\bf a}\cdot{\bf
n}\,dS+\int\limits_{\Gamma_2}{\bf a}\cdot{\bf n}\,dS=0,
$$
then by Lemma 1 there exists a divergence free  extension ${\bf
A}\in  W^{1,2}(\Omega)$ of ${\bf a}$ and there holds estimate (7).
Using this fact and standard results (see, e.g. \cite{Lad}) we can
find a weak solution ${\bf U}\in W^{1,2}(\Omega)$ of the Stokes
problem such that ${\bf U}-{\bf A}\in H(\Omega)$ and
$$
\nu\int\limits_\Omega\nabla{\bf U}\cdot\nabla\bfeta\,dx= 0
\quad\forall\;\bfeta\in H(\Omega). \eqno(19)
$$
Moreover,
$$
\|{\bf U}\|_{W^{1,2}(\Omega)}\leq c\|{\bf
a}\|_{W^{1/2,2}(\partial\Omega)}.\eqno(20)
$$

By a {\it weak solution} of problem (1) we understand a function
${\bf u}$ such that ${\bf w}={\bf u}-{\bf A}\in H(\Omega)$ and
satisfies  the integral identity
$$
\nu\int\limits_\Omega\nabla{\bf
w}\cdot\nabla\bfeta\,dx-\int\limits_\Omega\big(({\bf w}+{\bf
U})\cdot\nabla\big)\bfeta\cdot{\bf
w}\,dx-\int\limits_\Omega\big({\bf w}\cdot\nabla\big)\bfeta\cdot{\bf
U}\,dx
$$
$$
=\int\limits_\Omega\big({\bf U}\cdot\nabla\big)\bfeta\cdot{\bf
U}\,dx \qquad\forall\bfeta\in H(\Omega).\eqno(21)
$$

We shall prove the following
\\

{\bf Theorem 1. }  {\it Assume that  ${\bf a}\in
W^{1/2,2}(\partial\Omega)$ and let  condition $(6)$ be fulfilled. If
$\F=\int\limits_{\Gamma_2}{\bf a}\cdot{\bf n}\,dS\ge 0$, then
problem $(1)$ admits at least one weak solution.}
\\

{\bf Proof.} 1. We follow a contradiction  argument of J. Leray
\cite{Leray}. Although, this argument was used also in many other
papers (e.g. \cite{Lad1}, \cite{Lad}, \cite{KaPi1}, \cite{Amick}),
we reproduce here, for the  reader convenience, some details of it.
It is well known (e.g. \cite{Lad}) that  integral identity (21) is
equivalent to an operator equation in the space $H(\Omega)$ with a
compact operator, and, therefore, in virtue of the Leray--Schauder
fixed--point theorem, to prove the existence of a weak solution  to
Navier--Stokes problem (1)  it is sufficient to show that all
possible solutions of the integral identity
$$
\nu\int\limits_\Omega\nabla{\bf
w}\cdot\nabla\bfeta\,dx-\lambda\int\limits_\Omega\big(({\bf w}+{\bf
U})\cdot\nabla\big)\bfeta\cdot{\bf
w}\,dx-\lambda\int\limits_\Omega\big({\bf
w}\cdot\nabla\big)\bfeta\cdot{\bf U}\,dx
$$
$$
=\lambda\int\limits_\Omega\big({\bf
U}\cdot\nabla\big)\bfeta\cdot{\bf U}\,dx \qquad \forall\;\bfeta\in
H(\Omega)\eqno(22)
$$
are uniformly  bounded (with respect to $\lambda\in[0,\nu^{-1}]$) in
$H(\Omega)$. Assume this is false. Then there exist sequences $
\{\lambda_k\}_{k\in{\Bbb N}}\subset [0, \nu^{-1}]$ and $\{{\bf
w}_k\}_{k\in{\Bbb N}}\in H(\Omega)$ such that
$$
\nu\int\limits_\Omega\nabla{\bf
w}_k\cdot\nabla\bfeta\,dx-\lambda_k\int\limits_\Omega\big(({\bf
w}_k+{\bf U})\cdot\nabla\big)\bfeta\cdot{\bf
w}_k\,dx-\lambda_k\int\limits_\Omega\big({\bf
w}_k\cdot\nabla\big)\bfeta\cdot{\bf U}\,dx
$$
$$
=\lambda_k\int\limits_\Omega\big({\bf
U}\cdot\nabla\big)\bfeta\cdot{\bf U}\,dx\qquad\forall\,\bfeta\in
H(\Omega), \eqno(23)
$$
and
$$
\lim\limits_{k\to\infty}\lambda_k=\lambda_0\in[0, \nu^{-1}],\quad
\lim\limits_{k\to\infty}J_k=\lim\limits_{k\to\infty}\|{\bf
w}_k\|_{H(\Omega)}=\infty.\eqno(24)
$$
Let us take in (23) $\bfeta=J_k^{-2}{\bf w}_k$ and denote
$\widehat{\bf w}_k=J_k^{-1}{\bf w}_k$. Since
$$
\int\limits_\Omega\big(({\bf w}_k+{\bf U})\cdot\nabla\big){\bf
w}_k\cdot{\bf w}_k\,dx=0,
$$
we get
$$
\nu\int\limits_\Omega|\nabla\widehat{\bf
w}_k|^2\,dx=\lambda_k\int\limits_\Omega\big(\widehat{\bf
w}_k\cdot\nabla\big)\widehat{\bf w}_k\cdot{\bf U}\,dx+
J_k^{-1}\lambda_k\int\limits_\Omega\big({\bf
U}\cdot\nabla\big)\widehat{\bf w}_k\cdot{\bf U}\,dx. \eqno(25)
$$
Since $\|\widehat{\bf w}_k\|_{H(\Omega)}=1$, there exists a
subsequence $\{\widehat{\bf w}_{k_l}\}$   converging weakly
in $H(\Omega)$ to a vector field $\widehat{\bf w}\in H(\Omega)$.
Because of the compact imbedding
$$
H(\Omega)\hookrightarrow L^r(\Omega) \quad
\forall\,r\in(1,\infty),
$$
the subsequence $\{\widehat{\bf w}_{k_l}\}$ converges strongly in
$L^r (\Omega)$. Therefore, we can pass to a limit as $k_l\to\infty$
in equality (25). As a result we obtain
$$
\nu=\lambda_0\int\limits_\Omega\big(\widehat{\bf
w}\cdot\nabla\big)\widehat{\bf w}\cdot{\bf U}\,dx.\eqno(26)
$$

2. Let us return to integral identity (23). Consider the functional
$$
R_k(\bfeta)=\int\limits_\Omega\Big(\nu\nabla{\bf
w}_k\cdot\nabla\bfeta-\lambda_k\big(({\bf w}_k+{\bf
U})\cdot\nabla\big)\bfeta\cdot{\bf w}_k-\lambda_k\big({\bf
w}_k\cdot\nabla\big)\bfeta\cdot{\bf U}\Big)\,dx
$$
$$
-\lambda_k\int\limits_\Omega\big({\bf
U}\cdot\nabla\big)\bfeta\cdot{\bf U}\,dx \qquad \forall
\;\bfeta\in\wotwo(\Omega).
$$
Obviously, $R_k(\bfeta)$ is a linear functional, and  using (20) and
the imbedding theorem, we obtain
$$
\big|R_k(\bfeta)\big|\leq c\Big(\|{\bf w}_k\|_{H(\Omega)}+\|{\bf
w}_k\|_{H(\Omega)}^2
+\|{\bf a}\|_{W^{1/2,2}(\partial\Omega)}^2\Big)
\|\bfeta\|_{H(\Omega)},
$$
with constant $c$ independent of $k$. It follows from (23) that
$$
R_k(\bfeta)=0\qquad\forall\;\bfeta\in H(\Omega).
$$
Therefore, by Lemma 2, there exist  functions $p_k\in \widehat
L^2(\Omega)=\{q\in L^2(\Omega):\; \int\limits_\Omega q(x)\,dx=0\}$
such that
$$
R_k(\bfeta)=\int\limits_\Omega p_k\div
\bfeta\,dx\qquad\forall\;\bfeta\in \wotwo(\Omega),
$$
and
$$
\|p_k\|_{L^2(\Omega)}\leq c\Big(\|{\bf w}_k\|_{H(\Omega)}+\|{\bf
w}_k\|_{H(\Omega)}^2
 + \|{\bf a}\|_{W^{1/2,2}(\partial\Omega)}^2\Big). \eqno(27)
$$
The pair $\big({\bf w}_k, p_k)$  satisfies the integral identity
$$
\nu\int\limits_\Omega\nabla{\bf
w}_k\cdot\nabla\bfeta\,dx-\lambda_k\int\limits_\Omega\big(({\bf
w}_k+{\bf U})\cdot\nabla\big)\bfeta\cdot{\bf
w}_k\,dx-\lambda_k\int\limits_\Omega\big({\bf
w}_k\cdot\nabla\big)\bfeta\cdot{\bf U}\,dx
$$
$$
-\lambda_k\int\limits_\Omega\big({\bf
U}\cdot\nabla\big)\bfeta\cdot{\bf U}\,dx =\int\limits_\Omega p_k\div
\bfeta\,dx\qquad\forall \;\bfeta\in\wotwo(\Omega).\eqno(28)
$$
Let ${\bf u}_k={\bf w}_k+{\bf U}$. Then identity (28) takes the form
(see (19))
$$
\nu\int\limits_\Omega\nabla{\bf
u}_k\cdot\nabla\bfeta\,dx-\int\limits_\Omega p_k\,{\rm
div}\,\bfeta\,dx =-\lambda_k\int\limits_\Omega({\bf
u}_k\cdot\nabla\big){\bf u}_k\cdot\bfeta\,dx \;\;
\forall\,\bfeta\in\wotwo(\Omega).
$$
Thus,  $\big({\bf u}_k, p_k)$ might  be considered as a weak
solution to the Stokes problem
\begin{displaymath}
\left\{\begin{array}{rcl}-\nu\Delta{\bf u}_k +\nabla p_k & = & {\bf
f}_k \qquad \;\hbox{\rm in }\;\;\Omega,
\\[4pt]
\div{\bf u}_k & = & 0 \;\qquad \;\hbox{\rm in }\;\;\Omega,
\\[4pt]
{\bf u}_k &  = & {\bf a}
 \qquad\;\, \hbox{\rm on }\;\;\partial\Omega,
\end{array}\right.
\end{displaymath}
with the right--hand side ${\bf f}_k=-\lambda_k\big({\bf
u}_k\cdot\nabla\big){\bf u}_k$. Obviously, ${\bf f}_k\in
L^s(\Omega)$ for $s\in(1,2)$ and
$$
\|{\bf f}_k\|_{L^s(\Omega)}\leq c\|\big({\bf
u}_k\cdot\nabla\big){\bf u}_k\|_{L^s(\Omega)} \leq c\|{\bf
u}_k\|_{L^{2s/(2-s)}(\Omega)}\|\nabla{\bf u}_k\|_{L^{2}(\Omega)}
$$
$$
\leq c \Big( \big(\|{\bf w}_k\|_{H(\Omega)}+\|{\bf
U}\|_{W^{1,2}(\Omega)}\big)^2\Big) \leq c\Big(\|{\bf
w}_k\|_{H(\Omega)}^2
+\|{\bf a}\|_{W^{1/2,2}(\partial\Omega)}^2\Big),
$$
where $c$ is independent of $k$. By well known local regularity
results for the Stokes system (see \cite{Lad}, \cite{Galdibook}) we
have ${\bf w}_k\in W^{2,s}_{loc}(\Omega)$, $p_k\in
W_{loc}^{1,s}(\Omega)$, and  the estimate
$$
\|{\bf
w}_k\|_{W^{2,s}(\Omega^\prime)}+\|p_k\|_{W^{1,s}(\Omega^\prime)}\leq
c\Big( \|{\bf f}_k\|_{L^s(\Omega)}+\|{\bf
u}_k\|_{W^{1,2}(\Omega)}+\|{p}_k\|_{L^2(\Omega)}\Big)
$$
$$
\leq c\Big(\|{\bf w}_k\|_{H(\Omega)}^2+\|{\bf w}_k\|_{H(\Omega)}
+\|{\bf a}\|_{W^{1/2,2}(\partial\Omega)} +\|{\bf
a}\|_{W^{1/2,2}(\partial\Omega)}^2\Big),  \eqno(29)
$$
holds, where $\Omega^\prime$ is arbitrary domain with
$\overline{\Omega}\,^\prime\subset\Omega$ and the constant $c$
depends on ${\rm dist}\,(\Omega^\prime,
\partial\Omega)$ but not on $k$.

Denote $\widehat p_k=J_k^{-2} p_k $. It follows from (27) and (29)
that
$$
\|\widehat p_k\|_{L^{2}(\Omega)}\leq const, \quad \|\widehat
p_k\|_{W^{1,s}(\Omega^\prime)}\leq const
$$
for any $\overline\Omega^\prime\subset\Omega$ and $s\in(1,2)$.
Hence, from $\{\widehat p_{k_l}\}$ can be extracted a subsequence,
still denoted by $\{\widehat p_{k_l}\}$, which converges weakly in
$\widehat L^2(\Omega)$ and $W^{1,s}_{loc}(\Omega)$ to some function
$\widehat p\in W_{loc}^{1,s}(\Omega)\cap \widehat L^2(\Omega)$. Let
$\bfphi\in C_0^\infty(\Omega)$. Taking in (28)
$\bfeta=J_{k_l}^{-2}\bfphi$ and letting $k_l\to\infty$ yields
$$
-\lambda_0\int\limits_\Omega\big(\widehat{\bf
w}\cdot\nabla\big)\bfphi\cdot\widehat{\bf
w}\,dx=\int_\Omega\widehat p\,{\rm div}\varphi\,dx
\quad\forall\bfphi\in C^\infty_0(\Omega).
$$
Integrating by parts in the last equality, we derive
$$
\lambda_0\int\limits_\Omega\big(\widehat{\bf
w}\cdot\nabla\big)\widehat{\bf
w}\cdot\bfphi\,dx=-\int_\Omega\nabla\widehat p\cdot\varphi\,dx
\quad\forall\bfphi\in C^\infty_0(\Omega).\eqno(30)
$$
Hence, the pair $\big(\widehat{\bf w}, \widehat p\big)$ satisfies
for  almost all  $x\in\Omega$ the Euler equations
\begin{displaymath}
\left\{\begin{array}{rcl} \lambda_0\big(\widehat{\bf
w}\cdot\nabla\big)\widehat{\bf w}+\nabla\widehat p & = & 0,\\[4pt]
{\rm div }\,\widehat{\bf w} & = & 0,
\end{array}\right. \eqno(31)
\end{displaymath}
and $ \widehat{\bf w}\big|_{\partial\Omega}=0$. By Lemmas 6 and 7,
$\widehat p\in C(\overline{\Omega})\cap W^{1,2}(\Omega)$ and the
pressure $\widehat p(x)$ is constant on $\Gamma_1$ and $\Gamma_2$.

 Denote by $\widehat p_1$ and
$\widehat p_2$ values of $\widehat p(x)$ on $\Gamma_1$ and
$\Gamma_2$, respectively. Multiplying equations (31) by ${\bf U}$
and integrating by parts, we derive
$$
 \lambda_0\int\limits_\Omega\big(\widehat{\bf
w}\cdot\nabla\big)\widehat{\bf w}\cdot {\bf
U}\,dx=-\int\limits_\Omega\nabla\widehat p\cdot{\bf
U}\,dx=-\int\limits_{\partial\Omega}\widehat p \,{\bf a}\cdot{\bf
n}\,dS
$$
$$
= -\widehat p_1\int\limits_{\Gamma_1} {\bf a}\cdot{\bf
n}\,dS-\widehat p_2\int\limits_{\Gamma_2}{\bf a}\cdot{\bf
n}\,dS={\F}(\widehat p_1-\widehat p_2)\eqno(32)
$$
(see formula (6)). If either ${\F}=0$ or $\widehat p_1=\widehat
p_2$, it follows from (32) that
$$
\lambda_0\int\limits_\Omega\big(\widehat{\bf
w}\cdot\nabla\big)\widehat{\bf w}\cdot {\bf U}\,dx=0.\eqno(33)
$$
The last relation contradicts equality (26). Therefore, the norms
$\|{\bf w}\|_{H(\Omega)}$ of all possible solutions to identity (22)
are uniformly bounded with respect to $\lambda\in[0,\nu^{-1}]$ and
by Leray--Schauder fixed--point theorem problem (1) admits at least
one weak solution ${\bf u}\in
W^{1,2}(\Omega)$.\\

3. Up to this point our arguments were standard and followed those
of Leray \cite{Leray} (see also  \cite{KaPi1} and \cite{Amick}).
However, by the our assumptions  ${\F}>0$ and, in general, $\widehat
p_2\neq\widehat p_1$ (see a counterexample in \cite{Amick}). Thus,
(33) may be false. In order to prove that $\widehat p_1$ and
$\widehat p_2$ do coincide in the case ${\F}>0$, we use the property
of $\big(\widehat{\bf w},\widehat p\big)$ to be a limit (in some
sense) of solutions to the Navier--Stokes equations. Note that the
possibility of using this fact was already pointed up by Amick
\cite{Amick}.

 Let $\Phi_{k_l}=p_{k_l}+\dfrac{\lambda_{k_l}}{2}|{\bf
u}_{k_l}|^2$, where ${\bf u}_{k_l}={\bf w}_{k_l}+{\bf U} $,  be a
total head pressures corresponding to the solutions $\big({\bf
w}_{k_l}, p_{k_l}\big)$ of identities (25). Then $\Phi_{k_l}\in
W^{2,s}_{loc}(\Omega),\; s\in (1,2)$, satisfy almost everywhere in
$\Omega$ the equations
$$
\nu \Delta\Phi_{k_l}-\lambda_{k_l}\big({\bf
u}_{k_l}\cdot\nabla\big)\Phi_{k_l}=\nu \Big(\frac{\partial
{u}_{1k_l}}{\partial x_2}-\frac{\partial {u}_{2k_l}}{\partial
x_1}\Big)^2.
$$
It is well known \cite{GilbTrud1}, \cite{GilbTrud2} (see also
\cite{Mi2}) that for $\Phi_{k_l}$  one-side maximum principle holds
locally (since the boundary is only Lipschitz, $\Phi_{k_l}$ do not
have second derivatives up to the boundary). Set $\widehat
\Phi_{k_l}=J_{k_l}^{-2}\Phi_{k_l}$. It follows from (27), (29) that
the sequence
 $\widehat \Phi_{k_l}$ weakly converges   to $\widehat
\Phi=\widehat p+\dfrac{\lambda_0}{2}|\widehat{\bf  u}|^2$ in
$L^2(\Omega)\cap W^{1,s}_{loc}(\Omega),\, s\in (1,2)$. Therefore, by
Lemma 8, $\widehat \Phi$ satisfies the weak  one-sided maximum
principle and
$$
{\mathop{\hbox{\rm ess}\,\hbox{\rm
sup}}_{x\in\Omega}}\,\widehat\Phi(x)\leq {\mathop{\hbox{\rm
ess}\,\hbox{\rm
sup}}_{x\in\partial\Omega}}\,\widehat\Phi(x)=\max\{\widehat
p_1,\widehat p_2\}\eqno(34)
$$
(see (18)).

We conclude from equalities (26) and (32)
$$
(\widehat p_1-\widehat p_2){\F}=\nu >0.
$$
So, if ${\F}>0$, then
$$
\widehat p_2 < \widehat p_1.\eqno(35)
$$
Now,  it follows from (34), (35) that
$$
\int\limits_\Omega\widehat\Phi(x)\,dx \leq {\mathop{\hbox{\rm
ess}\,\hbox{\rm sup}}_{x\in\Omega}}\,\widehat\Phi(x)|{\Omega}|\leq
\widehat p_1|\Omega|,\eqno(36)
$$
where $|\Omega|$ means the measure of $\Omega$.\\

On the other hand, from  equation $(31_1)$ we obtain the identity
$$
0=x\cdot\nabla \widehat p(x)+\lambda_0 x\cdot\big(\widehat {\bf
w}(x)\cdot\nabla \big)\widehat{\bf w}(x)=\div\big[x\,\widehat p(x)+
\lambda_0 \big(\widehat {\bf w}(x)\cdot x\big)\widehat {\bf
w}(x)\big]
$$
$$
-\widehat p(x)\,\div x-\lambda_0 |\widehat {\bf
w}(x)|^2=\div\big[x\,\widehat p(x)+ \lambda_0 \big(\widehat {\bf
w}(x)\cdot x\big)\widehat {\bf w}(x)\big]-2\widehat\Phi(x).
$$
Integrating this identity we derive
$$
2\int\limits_\Omega\widehat\Phi(x)\,dx=\int\limits_{\partial\Omega}\widehat
p(x) \big(x\cdot{\bf n}\big)\,dS=\widehat p_1\int\limits_{\Gamma_1}
\big(x\cdot{\bf n}\big)\,dS+\widehat
p_2\int\limits_{\Gamma_2}\big(x\cdot{\bf n}\big)\,dS
$$
$$
=\widehat p_1\int\limits_{\Omega_1}{\rm div}\,x\,dx-\widehat
p_2\int\limits_{\Omega_2}{\rm div}\,x\,dx=2\big(\widehat
p_1|\Omega_1|-\widehat p_2|\Omega_2|\big).
$$
Hence,
$$
\int\limits_\Omega\widehat\Phi(x)\,dx=\widehat
p_1|\Omega_1|-\widehat p_2|\Omega_2|=\widehat
p_1|\Omega|+\big(\widehat p_1-\widehat p_2\big)|\Omega_2|.\eqno(37)
$$
Inequalities (36) and (37) yield
$$
\widehat p_1\leq\widehat p_2.
$$
This contradicts inequality (35). Thus, all solutions of integral
identity (22) are  uniformly bounded in $H(\Omega)$  and by the
Leray--Schauder fixed--point theorem there exists at least one weak
solution of problem (1). $\qquad\qquad\qquad\quad\Box$
\\
\\
{\bf Remark 2.} Let $\Omega=\{x:1<|x|<2\}$ be the annulus  and let
$(r,\theta)$ be polar coordinates in $\sR^2$.  If $f\in
C^\infty_0(1,2)$, then the pair $\widehat{\bf
w}=\big(\widehat{w}_r,\widehat{w}_\theta\big)$ and $\widehat p$ with
$$
\widehat{w}_r(r,\theta)=0,\;\; \widehat{w}_\theta(r,\theta)= f(r),
\quad \widehat
p(r,\theta)=\lambda_0\int_1^{r}\frac{f^2(t)}{t}\,dt\eqno(38)
$$
satisfy  both equations (31) and  the boundary condition
$\widehat{\bf w}\big|_{\partial\Omega}=0$ ($\widehat{w}_r$ and
$\widehat{w}_\theta$ are components of the velocity field in polar
coordinate system). On the other hand,
$$
0=\widehat p(x)\big|_{r=1}\neq \widehat
p(x)\big|_{r=2}=\lambda_0\int\limits _1^{2}\frac{f^2(t)}{t}\,dt>0.
$$
This simple example, due to Ch.J. Amick \cite{Amick} (see also
\cite{Galdi1}, v. II, p. 59), shows that, in general, the pressure
$\widehat p$ corresponding to the solution of Euler equations (31)
could have not equal constant values on different components of the
boundary.

It is interesting to observe that   for the  solution  like (38)
necessarily  holds $\widehat p_1>\widehat p_2$.  Indeed, writing the
Euler equations (31) in polar coordinates and integrating over
$\Omega$ yields
$$
\lambda_0\int_\Omega\frac{\widehat
{w}_\theta^2(r)}{r}\,drd\theta=\lambda_0\int_\Omega\frac{f^2(r)}{
r}\,drd\theta=\int_\Omega\frac{\partial p(r)}{\partial r}\,drd\theta
=\widehat p_1-\widehat p_2> 0.\eqno(39)
$$
The solution (38) cannot be a limit  of solutions to Navier--Stokes
problem (in the sense described in the proof of Theorem 1). If it is
so, then we conclude  from (26), (32) and (39) that $\F>0$. But
this, as it is proved in Theorem 1, leads to a contradiction.
\\

We emphasize that  in the case when $\F<0$ (inflow condition)
problem (1) remains unsolved. However, in this case  we do not know
any counterexample showing that for the solution of Euler equations
(31) the inequality $\widehat p_2>\widehat p_1$ holds.
\\

It is well known (see \cite{BOPI}, \cite{Galdi1}) that independently
of the sign of $\F$ problem (1) has a solution, if $|\F|$ is
sufficiently small. Using this result Theorem 1 can
be strengthened as follows \\

{\bf Theorem 2.} {\it Assume that  ${\bf a}\in
W^{1/2,2}(\partial\Omega)$ and let  condition $(6)$ be fulfilled.
Then there exists $\F_0>0$ such that for any $\F\in(-\F_0,+\infty)$
problem $(1)$ admits at least one weak solution.}

\end{document}